\documentclass{PoS}
\usepackage{amsmath}
\usepackage{amsthm}
\usepackage{amssymb}
\usepackage{amsfonts}
\usepackage{graphicx}
\usepackage{bm}
\usepackage{epsfig}

\usepackage{graphicx}
\usepackage{float}
\usepackage{wrapfig}

\newcommand{\eins}{\leavevmode\hbox{\small1\kern-3.8pt\normalsize1}}
\newcommand{\be}{\begin{eqnarray}}
\newcommand{\ee}{\end{eqnarray}}
\newcommand{\beq}{\begin{equation}}
\newcommand{\eeq}{\end{equation}}
\newcommand{\bqa}{\begin{eqnarray}}
\newcommand{\eqa}{\end{eqnarray}}

\title{Pion condensation and QCD phase diagram at finite isospin density}

\ShortTitle{Pion condensation and QCD phase diagram at finite isospin density}

\author{\speaker{Jens O. Andersen}\thanks{A footnote may follow.}\\
Department of Physics, Faculty of Natural Sciences, NTNU, 
Norwegian University of Science and Technology, H{\o}gskoleringen 5,
N-7491 Trondheim, Norway \\
        E-mail: \email{jensoa@ntnu.no}}

\author{Prabal Adhikari\\
St. Olaf College, Physics Department, 1520 St. Olaf Avenue,
Northfield, MN 55057, USA\\
E-mail:\email{adhika1@stolaf.edu}}

\author{Patrick Kneschke\\
Faculty of Science and Technology, University of Stavanger,
N-4036 Stavanger, Norway\\
E-mail:\email{patrick.kneschke@uis.no}}

\abstract{
  We use the Polyakov-loop extended two-flavor quark-meson model as 
a low-energy effective model for QCD to
study 1) the possibility of inhomogeneous chiral condensates and its
competition with a homogeneous
pion condensate in the $\mu$--$\mu_I$ plane at $T=0$ and
2) the phase diagram in the $\mu_I$--$T$ plane.
In the $\mu$--$\mu_I$ plane,
we find that an inhomogeneous chiral condensate only exists for
pion masses lower that 37.1 MeV and does not coexist with a
homogeneous pion condensate.
In the $\mu_I$--$T$ plane, we find 
that the phase transition to a Bose-condensed phase is of second order
for all values of $\mu_I$ and we find that there is no pion condensation
for temperatures larger than approximately 187 MeV. 
The chiral critical line joins the critical line
for pion condensation at a  point, whose position depends on the
Polyakov-loop potential and the sigma mass.
For larger values of $\mu_I$ these curves are on top of each other.
The deconfinement line enters smoothly the phase with
the broken $O(2)$ symmetry.
We compare our results with recent lattice simulations and find overall
good agreement.}

\FullConference{XIII Quark Confinement and the Hadron Spectrum - Confinement2018\\
		31 July - 6 August 2018\\
		Maynooth University, Ireland}

\begin{document}

\begin{figure}[htb]
\centering{\includegraphics[scale=0.41]{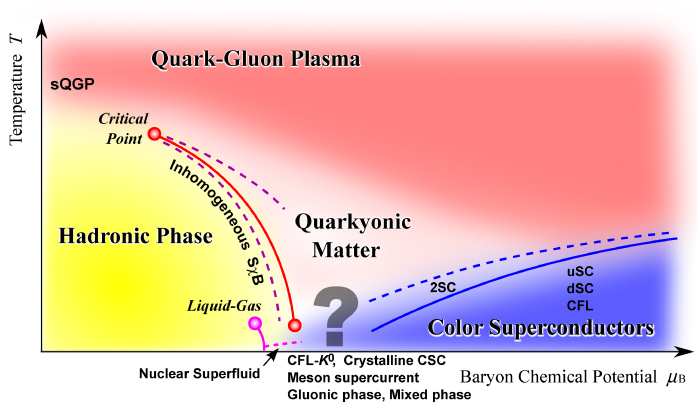}} 
\caption{QCD phase diagram in the $\mu$-$T$ plane. Figure is taken from
Ref.~\cite{kenji2}.}
\label{diag}
\end{figure}

Fig.~\ref{diag} shows the (conjectured) QCD phase diagram in the $\mu$-$T$
plane. 
I say conjectured, because
only a few exact results exist. For example, at asympotically high temperatures,
we know due to asymptotic freedom
that there is a weakly interacting quark-gluon plasma.
Likewise, at asymptotically large densities, we have the color-flavor-locked
phase. This is a color-superconducting phase, whose existence is guaranteed
by an attractive channel via one-gluon exchange.
A severe problem that arises as one tries to map out the phase diagram,
is that one cannot use lattice simulations at
large baryon chemical potentials due to the sign problem. One must therefore
resort to models. The model dependence has been indicated by a question mark
in the figure.
Two popular models are the Nambu-Jona-Lasinio (NJL)
and the quark-meson (QM) models, often extended by coupling it to
the Polyakov loop in order to account for confinement.

We can add more external parameters or axes to Fig.~\ref{diag}
to construct a multidimensional space.
For example, 
instead of having a common quark chemical potential for all
flavors, we can introduce a chemical potential $\mu_f$ for each of them.
For $n_f=2$, we can therefore either use $\mu_u$ and $\mu_d$ or equivalently
use baryon and isospin chemical potentials $\mu_B$ and $\mu_I$.
The relations between the
two sets of chemical potentials are
$\mu_u=\mu+\mu_I$ and $\mu_d=\mu-\mu_I$
where $\mu={1\over3}\mu_B$ is the quark chemical potential.
A nonzero value of $\mu_I$ introduces an
imbalance between the $u$ and the $d$-quarks, which gives rise to
Bose-Einstein condensation of pions if it is large enough. 

\begin{figure}[htb]
\centering{ \includegraphics[scale=2.4]{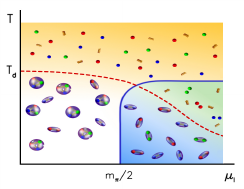}} 
 \caption{Phase diagram in the $\mu_I$-$T$ plane. Figure is taken from
   \cite{gergy3}.}
\label{pdg2}
\end{figure}

The phase diagram in the $\mu_I$-$T$ plane was conjectured in 2001 by
Son and Stephanov~\cite{son}, and a possible scenario is
shown in Fig.~\ref{pdg2}.
In the lower left corner, there is a
hadronic phase. As one cranks up the temperatue, there is a deconfinement
transition to a quark-gluon plasma phase. This transition is
indicated by the dashed red line.
Along the $\mu_I$-axis, one enters
a Bose-condensed phase of pions. This phase breaks an $O(2)$ symmetry
associated with the conservation of the third component of
isospin.
At $T=0$, the onset of
charged pion condesation is exactly at
$\mu_I={1\over2}m_{\pi}$.~\footnote{Another common definition of
$\mu_I$ differs by a factor of
two such that the onset is at $\mu_I=m_{\pi}$.} For larger values yet of
$\mu_I$, there is a crossover transition to a BCS
state.~\footnote{The BEC and BCS phases break the same symmetries so there
is no phase transition but rather an analytic crossover~\cite{son}.}
In this phase, there is a condensate
of weakly bound Cooper pairs, rather than a condensate of tightly bound
charged pions. The blue line indicates the transition to a Bose-condensed
phase (blue region) or a BCS state (green region).

In this talk, I would like to discuss certain aspects of
the QCD phase diagram, namely that of pion condensation at finite
isospin density~\cite{patrick,prabal}. 
Specifically, I want to address the following 
\begin{enumerate}
\item 
  Phase diagram in the $\mu$-$\mu_I$ plane at $T=0$. Inhomogeneous phases and
  competition with an inhomogeneous pion condensate
\item Phase diagram in the $\mu_I$-$T$ plane. Pion condensation,
  BEC-BCS transition, chiral and deconfinement transitions.

\end{enumerate}

\section{Quark-meson model}
In order to map out the phase diagrams, we employ
the quark-meson model as 
an effective low-energy model of QCD.
The Minkowski Lagrangian including
$\mu_u$ and $\mu_d$ reads
\bqa \nonumber
{\cal L}&=&
{1\over2}\left[(\partial_{\mu}\sigma)(\partial^{\mu}\sigma)
+(\partial_{\mu} \pi_3)(\partial^{\mu} \pi_3)
\right]
+(\partial_{\mu}+2i\mu_I\delta_{\mu}^0)\pi^+(\partial^{\mu}-2i\mu_I\delta_{0}^{\mu})
\pi^-
\nonumber-{1\over2}m^2(\sigma^2+\pi_3^2
\\ &&
+2\pi^+\pi^-)
-{\lambda\over24}(\sigma^2+\pi_3^2+2\pi^+\pi^-)^2
+h\sigma+\bar{\psi}\left[
i/\!\!\!\!\partial
+\mu_f
\gamma^0
-g(\sigma+i\gamma^5{\boldsymbol\tau}\cdot{\boldsymbol\pi})\right]\psi\;.
\label{lag}
\eqa
We will in the following allow for an inhomogeneous chiral condensate,
several different ans\"atze have been discussed in the literature, for
example a chiral-density wave (CDW) and a chiral soliton
lattice.
We choose the simplest ansatz, namely that of a CDW,
\bqa\nonumber
\sigma=\phi_0\cos(qz)\;,\hspace{0.5cm}
\pi_1=\pi_0
\;,\hspace{0.5cm}
\pi_2=0
\;,\hspace{0.5cm}
\pi_3=\phi_0\sin(qz)\;,
\eqa
where $\phi_0$ is the magnitude of the chiral condesate, $q$ is a wavevector,
and $\pi_0$ is a homogeneous pion condensate.
Below we will express the effective potential using
the variables $\Delta=g\phi_0$ and $\rho=g\pi_0$.

There are two technical details I would like to
briefly mention, namely that of parameter fixing and regulator artefacts.
In mean-field calculations it is common to determine the
parameters of the Lagrangian at tree level. However, this is inconsistent.
One must determine the parameters to the same accuracy as one is calculating
the effective potential. We are calculating the effective potential
to one loop in the large-$N_c$ limit, which means that we integrate
over the fermions, but treat the mesons at tree level.
Consequently, we must determine the parameters in Eq. (\ref{lag})
in the same approximation. If one determines the
parameters at tree level and the effective potential in the one-loop
large-$N_c$ approximation, the onset of BEC will not be at
$\mu_I={1\over2}m_{\pi}$, in fact in some case there are substantial
deviations from this exact result. This point has been ignored in most
calculations to date.
Secondly,
in calculations using the NJL model, one is typically using a hard
momentum cutoff. In cases with inhomogeneous phases, this leads to an
asymmetry in the states included due to the sign of $q$ in the
different quark dispersion relations. This leads to a $q$-dependent
effective potential even in the limit $\Delta\rightarrow0$. This is
inconsistent and one must typically subtract a  $q$-dependent term
to remedy this.
In the present calculations in the QM model,
we are using dimensional regularization.
We have explicitly checked that the effective potential is consistent, i.e.
it is independent of the wavevector $q$ in the limit
$\Delta\rightarrow0$.

Let us return to the QM Lagrangian (\ref{lag}).
The quark energies are
\bqa\nonumber
  E_u^{\pm}&=&E(\pm q,-\mu_I)\;,
\hspace{0.5cm}
E_d^{\pm}=E(\pm q,\mu_I)\;,
\hspace{0.5cm}
E_{\bar{u}}^{\pm}=E(\pm q,\mu_I)\;,
\hspace{0.5cm}E_{\bar{d}}^{\pm}=E(\pm q,-\mu_I)\;, 
\eqa
where
\bqa
E(q,\mu_I)&=&
\left[
\left(\sqrt{p_{\perp}^2+
\left(\sqrt{p_{\parallel}^2+\Delta^2}+{q\over2}\right)^2}+
{\mu_I}\right)^2
+\rho^2\right]^{1\over2}\;.
\eqa
Note in particular that the quark energies depend on the isospin chemical
potential.
After regularization and renormalization, the zero-temperature part of the
effective potential
can be expressed in terms of the physical meson masses and the pion-decay
constant as~\footnote{We have been combining the on-shell and
  $\overline{\rm MS}$
schemes~\cite{patrick}.}
\bqa\nonumber
V_{\rm 1-loop}&=&
{1\over2}f_{\pi}^2q^2
\left\{1-\dfrac{4 m_q^2N_c}{(4\pi)^2f_\pi^2}
\left[\log\mbox{$\Delta^2+\rho^2\over m_q^2$}
  +F(m_\pi^2)+m_\pi^2F^{\prime}(m_\pi^2)
  \right]
\right\}{\Delta^2\over m_q^2}
\\ && \nonumber
+\dfrac{3}{4}m_\pi^2 f_\pi^2
\left\{1-\dfrac{4 m_q^2N_c}{(4\pi)^2f_\pi^2}m_\pi^2F^{\prime}(m_\pi^2)
\right\}\dfrac{\Delta^2+\rho^2}{m_q^2}
\\ \nonumber &&
 -\dfrac{1}{4}m_\sigma^2 f_\pi^2
\left\{
1 +\dfrac{4 m_q^2N_c}{(4\pi)^2f_\pi^2}
\left[ \left(1-\mbox{$4m_q^2\over m_\sigma^2$}
\right)F(m_\sigma^2)
 +\dfrac{4m_q^2}{m_\sigma^2}
-F(m_\pi^2)-m_\pi^2F^{\prime}(m_\pi^2)
\right]\right\}\dfrac{\Delta^2+\rho^2}{m_q^2} 
\\ \nonumber &&
-2\mu_I^2f_\pi^2
\left\{1-\dfrac{4 m_q^2N_c}{(4\pi)^2f_\pi^2}
\left[\log\mbox{$\Delta^2+\rho^2\over m_q^2$}
+F(m_\pi^2)+m_\pi^2F^{\prime}(m_\pi^2)\right]
\right\}{\rho^2\over m_q^2}
\\ \nonumber
 & & + \dfrac{1}{8}m_\sigma^2 f_\pi^2
\Bigg\{ 1 -\dfrac{4 m_q^2  N_c}{(4\pi)^2f_\pi^2}\Bigg[
\dfrac{4m_q^2}{m_\sigma^2}
\left( 
\log\mbox{$\Delta^2+\rho^2\over m_q^2$}
-\mbox{$3\over2$}
\right) -\left( 1 -\mbox{$4m_q^2\over m_\sigma^2$}\right)F(m_\sigma^2)
\\ &&\nonumber
+F(m_\pi^2)+m_\pi^2F^{\prime}(m_\pi^2)\Bigg]\Bigg\}
\dfrac{(\Delta^2+\rho^2)^2}{m_q^4}
- \dfrac{1}{8}m_\pi^2 f_\pi^2
\left[1-\dfrac{4 m_q^2N_c}{(4\pi)^2f_\pi^2}m_\pi^2F^{\prime}(m_\pi^2)\right]
\dfrac{(\Delta^2+\rho^2)^2}{m_q^4}
\\ &&
-m_\pi^2f_\pi^2\left[
1-\dfrac{4 m_q^2  N_c}{(4\pi)^2f_\pi^2}m_\pi^2F^{\prime}(m_\pi^2)
\right]\dfrac{\Delta}{m_q}\delta_{q,0}
+V_{\rm fin}\;,
\label{fullb}
\eqa
where  $V_{\rm fin}$ is a finite term that must be evaluated numerically.
The linear term is responsible for explicit
chiral symmetry breaking. It only contributes when $q=0$, i.e. in the
homogeneous case; for nonzero $q$ this term averages to zero over
a sufficiently large spatial volume of the system.
The finite-temperature part of the one-loop effective potential is
\bqa
V_{T}&=&-2N_cT\int_p\bigg\{
\log\Big[1+e^{-\beta(E_{u}-\mu)}\Big]
+\log\Big[1+e^{-\beta(E_{\bar{u}}+\mu)}\Big]
\bigg\}
+{u\leftrightarrow d}\;.
\label{fd}
\eqa
\section{Coupling to the Polyakov loop}
The Wilson line which wraps all the way around in imaginary time
is defined as
\bqa
L({\bf x})&=&{\cal P}\exp\left[
i\int_0^{\beta}d\tau A_4({\bf x},\tau)
\right]\;,
\label{ldef}
\eqa
where ${\cal P}$ is time ordering.
The Wilson line is not gauge invariant, but taking the trace of it
gives a gauge-invariant object, namely the Polyakov loop ${\rm Tr}\,L$.
The Polyakov loop is, however, not invariant under the socalled center
symmetry of the (pure glue) QCD Lagrangian, but transforms as 
${\rm Tr}\,L\rightarrow e^{i{n\over N_c}}{\rm Tr }\,L$, where $n=0,1,2...N_c-1$ and
$e^{i{n\over N_c}}$ is one of
the $N_c$  roots of unity. A nonzero expectation value of the
Polyakov loop therefore signals the breaking of the center symmetry.
The expectation value of the correlator of Polyakov loop and its conjugate
is related to the free energy of a quark and an antiquark.
Employing cluster decomposition, the expectation value of the Polyakov
loop itself is related to the free energy of a single quark;
A vanishing value of ${\rm Tr}\,L$ corresponds to an infinite free energy
of a single quark and a vanishing value of
${\rm Tr}\,L^{\dagger}$ corresponds to an infinite free energy
of an antiquark, and therefore signals confinement.
If we denote by $\Phi$ the expectation value of 
${\rm Tr}\,L$ and $\bar{\Phi}$ of ${\rm Tr}\,L^{\dagger}$,
the medium contribution reads
  \bqa\nonumber
V_T&=&-2T\int{d^3p\over(2\pi)^3}
\bigg\{
    {\rm Tr}\log\Big[1+3(\Phi+\bar{\Phi}e^{-\beta(E_u-\mu)})
      e^{-\beta(E_u-\mu)}+e^{-3\beta(E_u-\mu)}\Big]
\\ 
&&+{\rm Tr}\log\Big[1+3(\bar{\Phi}+\Phi e^{-\beta(E_{\bar{u}}+\mu)})
  e^{-\beta(E_{\bar{u}}+\mu)}
+e^{-3\beta(E_{\bar{u}}+\mu)}\Big]
\bigg\}
+{u\leftrightarrow d}
\;.
\eqa 
We note that the symmetry between $\Phi$ and $\bar{\Phi}$ implies
$\Phi=\bar{\Phi}$.
The expression for $V_T$ is also manifestly
real, reflecting that there is no sign problem at finite $\mu_I$ and zero
$\mu_B$. Finally, for $\Phi=\bar{\Phi}=1$, we recover the standard
expression for the fermionic contribution (\ref{fd}) for $N_c=3$.

We also need to introduce a potential ${\cal U}$ from the glue sector.
The terms are constructed from $\Phi$ and $\bar{\Phi}$ such that
it satisfies the symmetries. There are several such potentials
on the market. We choose the following logarithmic potential~\cite{ratti2}
\bqa
{{\cal U}\over {T^4}}&=&-{1\over2}a \Phi\bar{\Phi} + b
\log\left[ 1 - 6\Phi\bar{\Phi} +4(\Phi^3 +\bar{\Phi}^3) -3(\Phi\bar{\Phi})^2
\right]\;,
\label{pglog}
\eqa
with
$a = 3.51 -2.47\left({T_0\over T}\right) +15.2\left({T_0\over T}\right)^2$
and $b = -1.75\left({T_0\over T}\right)^3$\;.
The parameters are determined such that the potential reproduces the
pressure for pure-glue QCD as calculated on the lattice for temperatures
around the critical temperature. Since the critical temperature depends
on the number of flavors $n_f$ and $\mu_I$, one can refine
the potential by making $T_0$ dependent on these parameters,
$T_0(N_f,\mu_I)=T_{\tau}e^{-1/(\alpha_0b(\mu_I))}$ with
$b(\mu_I)={1\over6\pi}(11N_c-2N_f)-b_{\mu_I}{\mu_I^2\over T_{\tau}^2}$
and $T_{\tau}=1.77\;\text{GeV}$ and $\alpha_0 = 0.304$
are determined such that the transition temperature for pure glue at $\mu_I=0$
is $T_0=270$ MeV~\cite{pawlow}. 
In the numerical work, we use 
$m_{\pi}=140$ MeV, $m_{\sigma}=500$ MeV, $\Delta=300$ MeV, and $f_{\pi}=93$ MeV.


\section{Phase diagram in the $\mu$-$\mu_I$ plane}
\begin{figure}[htb]
\begin{center}
  \includegraphics[width=0.45\textwidth]{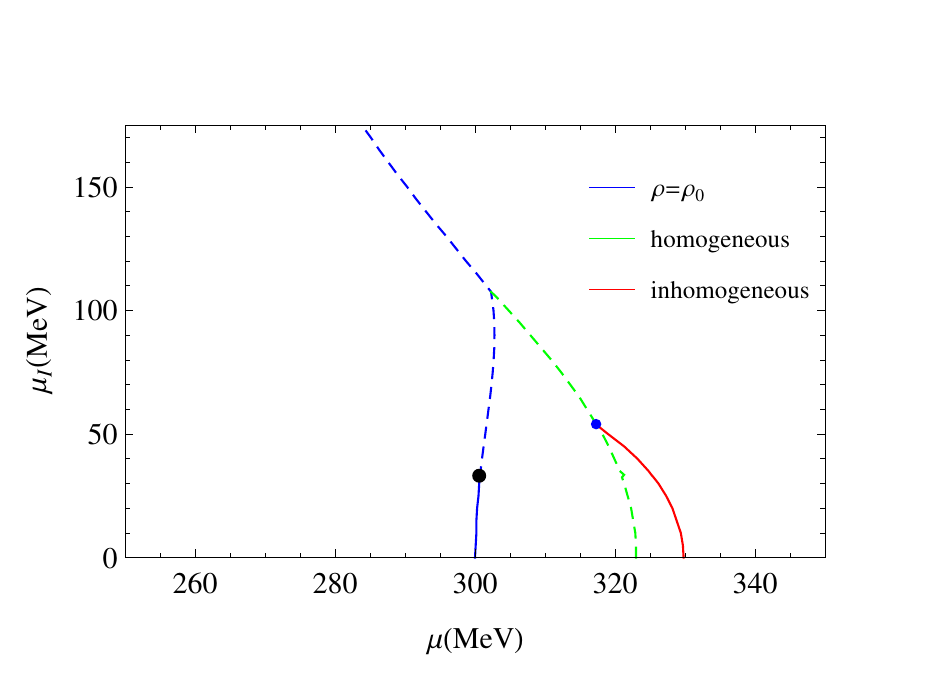}
\includegraphics[width=0.45\textwidth]{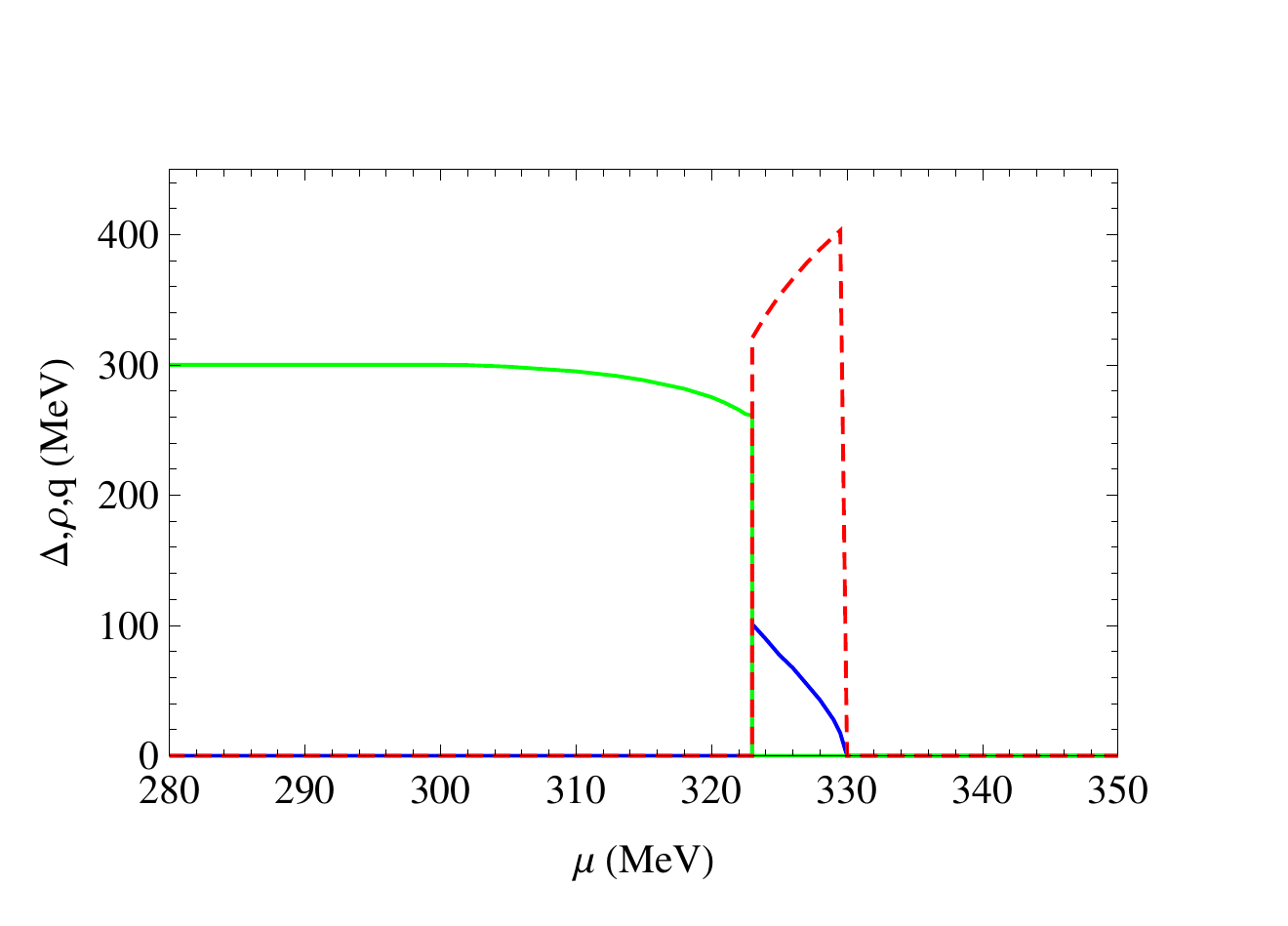}
\caption{Phase diagram in the chiral limit (left) and $\rho$, $\Delta$, and
  $q$ as functions of $\mu$ for $\mu_I=5$ MeV (right). See main text for
  details.}
\label{pionphase2}
\end{center}
\end{figure}

We first discuss the results for the phase diagram in the $\mu$-$\mu_I$ plane
at $T=0$  in the chiral limit, which
is shown in the left panel of Fig.~\ref{pionphase2}.
Dashed lines indicate first-order transitions while solid lines indicate
second-order transitions. The black dot indicates the
endpoint of the first-order line.
The vacuum phase is part of the $\mu$-axis (recall that in the chiral limit,
the onset of pion condensation is at $\mu_I=0$), ranging from $\mu=0$
to $\mu=300$ MeV. In this phase all the thermodynamic functions are independent
of the quark chemical potentials.
In the region to the left of the blue line, there is a nonzero pion condensate
which is independent of $\mu$.
In the wedge-like shaped region between the blue and green
lines, there is a phase with a $\mu$-dependent pion condensate and
a vanishing chiral condensate. In this phase the isospin and
quark densities are nonzero. In the region between the green and red lines,
we have an inhomogeneous phase with a nonzero wavevector $q$. In this
phase, the pion condensate is zero; thus there is no coexistence of
an inhomogeneous chiral condensate and a homogeneous pion condensate.
Finally,  the  region  to  the  right  of  red,  blue,
and green line segments is the symmetric phase, where
$\Delta=\rho=q=0$. The  blue  dot  marks  the  Lifshitz
point where the homogeneous, inhomogeneous and chi-
rally symmetric phases connect.
In the right panel of Fig.~\ref{pionphase2}, we show 
the pion condensate $\rho$ (green line),
the magnitude of the chiral condensate $\Delta$ (blue line), and the
wavevector $q$ (red line) as functions of $\mu$ for fixed $\mu_I=5$ MeV.
This corresponds to a horizontal line in the phase diagram.

\begin{figure}[htb]
\begin{center}
  \includegraphics[width=0.44\textwidth]{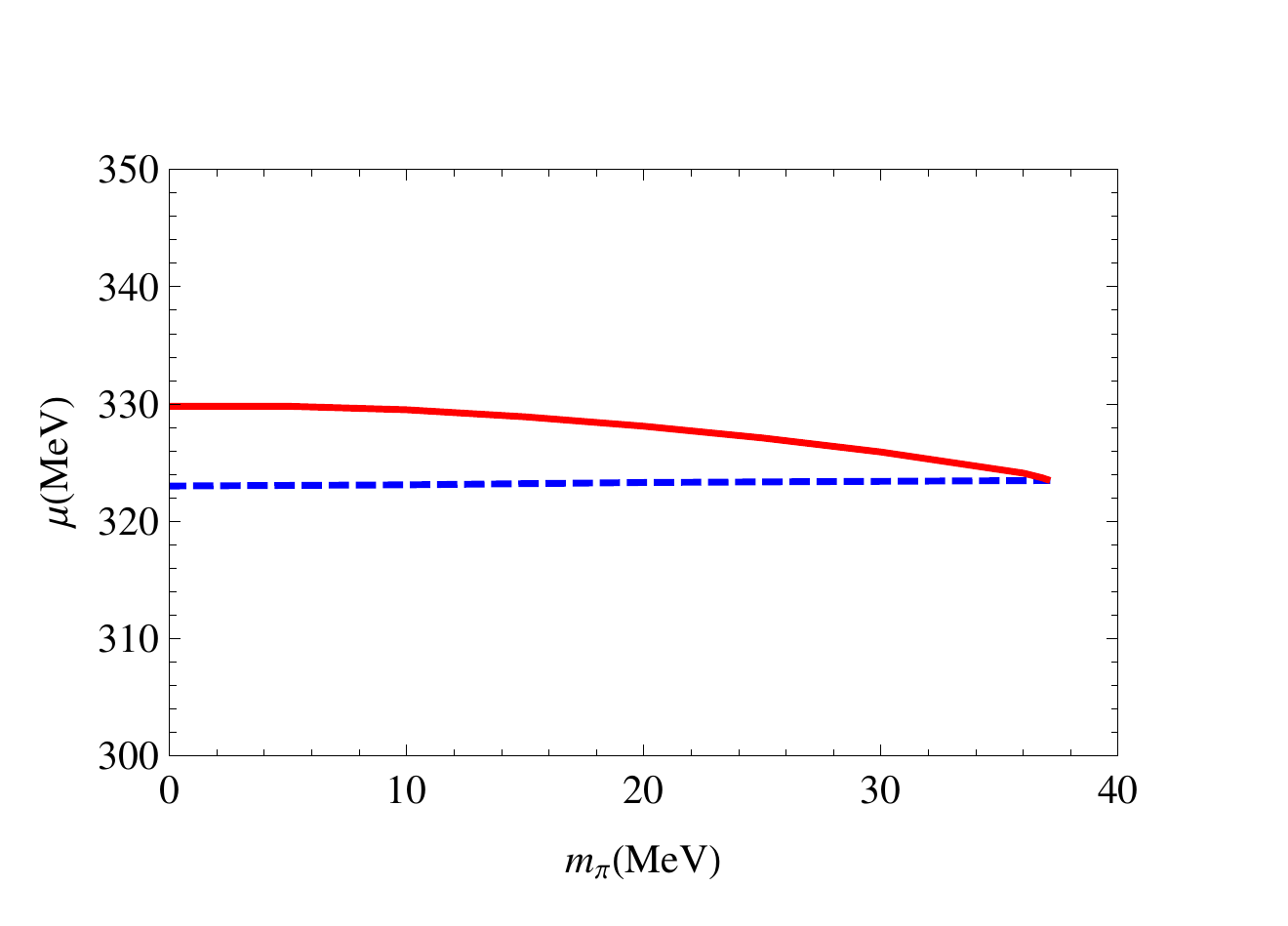}
  \includegraphics[width=0.44\textwidth]{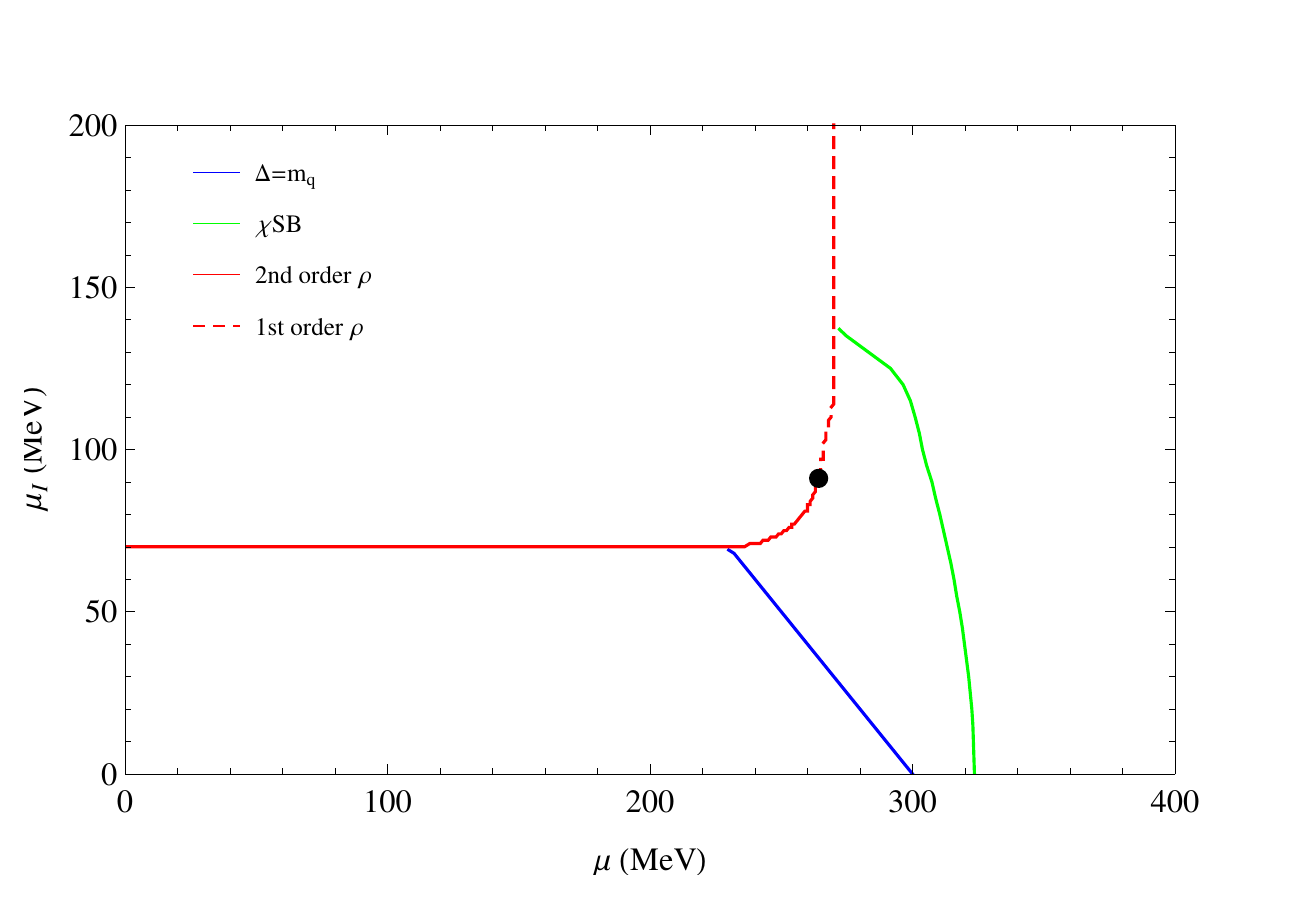}
\end{center}
\caption{Window of inhomogeneous phase as a function of the pion mass
  for $\mu_I=0$ (left) and the homogeneous phase diagram at the physical
point (right).}
\label{bondary}
\end{figure}

In the left panel of Fig.~\ref{bondary}, we show the
window of inhomogeneous chiral condensate for $\mu_I=0$, i.e. on the
$\mu$-axis as a function of the pion mass.
The inhomogeneous phase ceases to exist for pion masses larger than
$37.1$ MeV and so is not present at the physical point
This is contrast to the QM study in Ref.~\cite{nickel} where an
inhomogeneous phase is found at the physical point.
However, in that paper the parameters were
determined at tree level, which may explain the difference with
the results presented here.
The question of inhomogeneous phases in QCD was also addressed by
Buballa at this meeting using the NJL model, in particular
discussing the role of the quark masses~\cite{bubi}.
Performing a Ginzburg-Landau analysis, they find an inhomogeneous
phase which
shrinks with increasing quark mass, but survives at the physical point.

In the right panel of
Fig. \ref{bondary}, we show the phase diagram at the physical point.
The thermodynamic observables are independent
of $\mu$ and $\mu_I$ in the region bounded by the 
$\mu_I$ and $\mu$ axes, and the straight lines given by 
$\mu+\mu_I=gf_{\pi}=m_q$ (blue line) and $\mu_I=\mu_i^c={1\over2}m_{\pi}$
(red line).
We therefore refer to this as the vacuum phase.
The red line shows the phase boundary between a phase with
$\rho=0$ and a pion-condensed phase. The transition is second order
when the red line is solid and first order when it is dashed. The
solid dot indicates the position of the critical end point where the
first-order line ends, and is located at
$(\mu, \mu_I) = (264,91)$ MeV. The green line indicates the boundary
between a chirally broken phase and a phase where chiral symmetry 
is approximately restored. This line is defined by the inflection point
of the chiral condensate in the $\mu$-direction.
The region bounded by the three lines is a phase
with chiral symmetry breaking but no pion condensate. The 
effective potential depends on $\mu$ and $\mu_I$ and therefore
the quark and isospin densities are nonzero.

\section{Phase diagram in the $\mu_I$-$T$ plane}

\begin{figure}[!htb]
\begin{center}
  \includegraphics[width=0.45\textwidth]{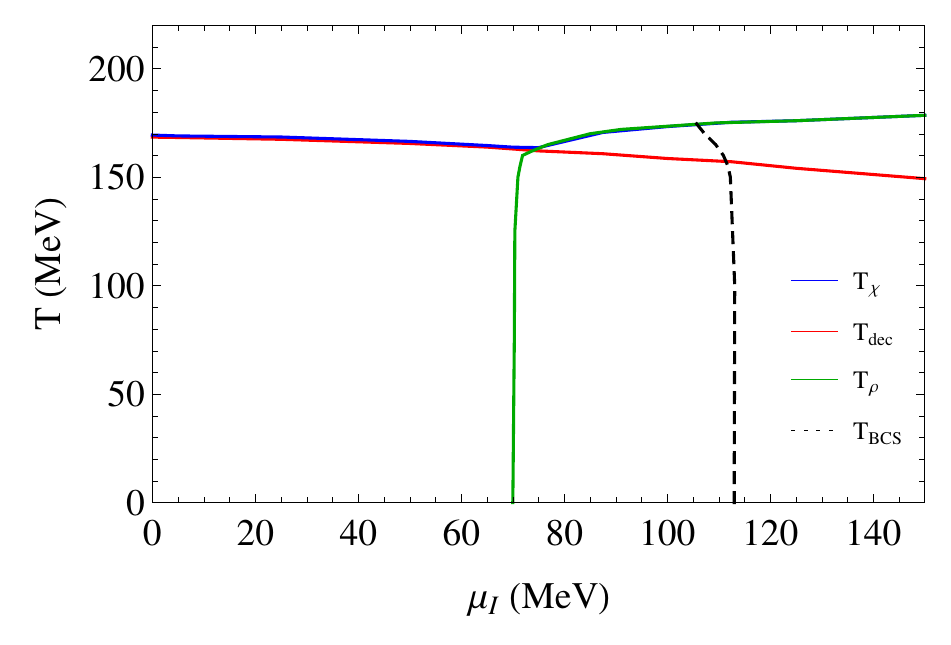}
  \includegraphics[width=0.45\textwidth]{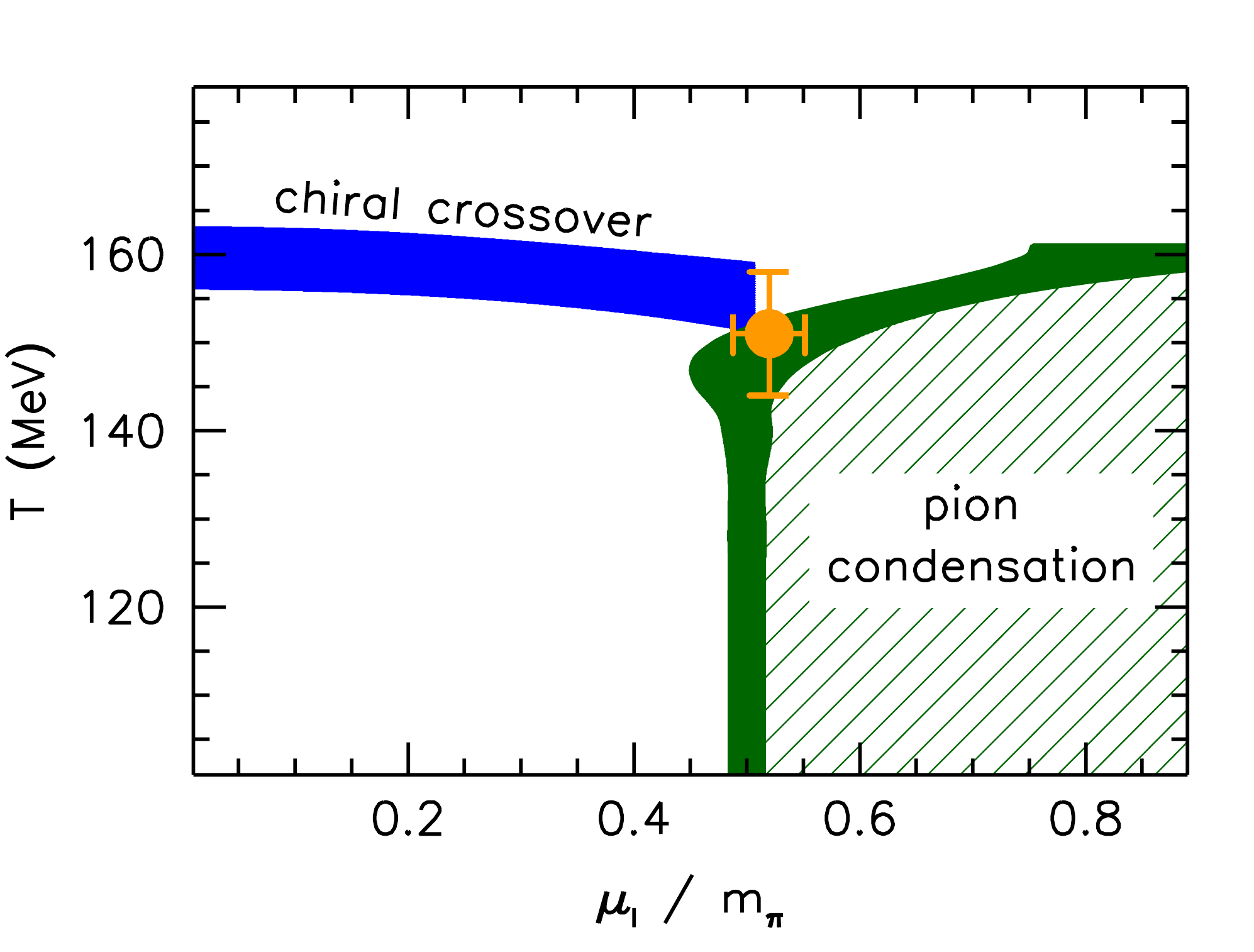}
  \end{center}
\caption{Phase diagram in the $\mu_I$-$T$ plane from the PQM model
(left) and from the lattice simulations of Refs.~\cite{gergy1,gergy2,gergy3}.}
\label{pionphase}
\end{figure}
In the left panel of Fig.~\ref{pionphase},
we show the phase diagram in the $\mu_I$-$T$ diagram
resulting from our PQM model calculation. The green line indicates
the transition to a pion-condensed phase, where the $O(2)$-symmetry associated
with the conservation of the third component of the isospin is broken.
This transition is second order everywhere. At $T=0$ this transition takes
place at $\mu_I={1\over2}m_{\pi}$ by construction.
The chiral transition line is in blue, while the transition line for the
deconfinement transition is in red.
In the noncondensed phase, these coincide.
When these lines meet the BEC line, 
they depart and the former coincides the BEC line.
The transition line for deconfinement penetrates into condensed phase.
Finally, the dashed line indicates the BEC-BCS crossover defined by
the condition $\Delta>\mu_I$, i.e. when the dispersion relations for the 
$u$-quark and $\bar{d}$-quark
no longer have their minima at $p=0$, but rather at $p=\sqrt{\Delta^2-\mu_I^2}$.
In the right panel of Fig.~\ref{pionphase}, we see the lattice results of
Refs.~\cite{gergy1,gergy2,gergy3}.
The blue band indicates the chiral crossover. 
Within the uncertainty it coincides with 
the deconfinement transition. The green band indicates
the second-order transition to a Bose-Einstein condensed state. The three
transition meet at the yellow point. Their simulations also indicate
that the deconfinement transition temperature
decreases and the transition line smoothly penetrates into the
pion-condensed phase.

\section{Summary}
I would like to finish this talk by highlighting our main results and
the comparison with lattice results from Refs.~\cite{gergy1,gergy2,gergy3}.

\begin{enumerate}
\item Rich phase diagrams. Inhomogeneous chiral condensate excluded
for $m_{\pi}>37.1$ MeV.

\item No inhomogeneous chiral condensate coexists with a
  homogeneous pion condensate.
\item Good agreement between lattice simulations and model calculations:
\begin{enumerate}
\item Second-order transition to a BEC state, which is in the
$O(2)$ universality class.
At $T=0$, onset of pion condensation at exactly $\mu_I^c={1\over2}m_{\pi}$.  
\item BEC and chiral transition lines merge at large $\mu_I$.
\item The deconfinement transition smoothly penetrates
into the BEC phase.
\end{enumerate}
\end{enumerate}


\end{document}